\documentclass[a4paper]{IEEEtran}
\usepackage{subfigure}
\usepackage{graphicx}
\usepackage{color}
\usepackage{amsmath}
\usepackage{epsfig}
\usepackage{amsfonts}
\usepackage{amssymb}
\usepackage{gensymb}
\usepackage{epstopdf}
\usepackage{cite}
\usepackage[top=2.8cm, left=1.5cm, right=1.5cm, bottom=1.8cm]{geometry}
\usepackage{algorithm}
\usepackage[noend]{algpseudocode}

\newcommand{\argmax}{\mathop{\text{argmax}}}
\newcommand{\argmin}{\mathop{\text{argmin}}}

\newcommand{\Define}{\triangleq}

\begin{document}
\title{{\LARGE
Multiuser Media-based Modulation for Massive MIMO Systems}}
\author{Bharath Shamasundar and A. Chockalingam \\
Department of ECE, Indian Institute of Science, Bangalore 560012 }
\maketitle

\thispagestyle{empty}
\begin{abstract}
In this paper, we consider {\em media-based modulation (MBM)}, an 
attractive modulation scheme which is getting increased research 
attention recently, for the uplink of a massive MIMO system. 
Each user is equipped with one transmit antenna with multiple radio 
frequency (RF) mirrors (parasitic elements) placed near it. The base 
station (BS) is equipped with tens to hundreds of receive antennas.
MBM with $m_{rf}$ RF mirrors and $n_r$ receive antennas over a 
multipath channel has been shown to asymptotically (as 
$m_{rf}\rightarrow \infty$) achieve the capacity of $n_r$ parallel 
AWGN channels. This suggests that MBM can be attractive for use in 
massive MIMO systems which typically employ a large number of receive 
antennas at the BS. In this paper, we investigate the potential 
performance advantage of multiuser MBM (MU-MBM) in a massive MIMO 
setting. Our results show that multiuser MBM (MU-MBM) can 
significantly outperform other modulation schemes. For example, 
a bit error performance achieved using 500 receive antennas at the 
BS in a massive MIMO system using conventional modulation can be
achieved using just 128 antennas using MU-MBM. Even multiuser
spatial modulation, and generalized spatial modulation in the same 
massive MIMO settings require more than 200 antennas to achieve the 
same bit error performance. Also, recognizing that the MU-MBM signal 
vectors are inherently sparse, we propose an efficient MU-MBM signal 
detection scheme that uses compressive sensing based reconstruction 
algorithms like orthogonal matching pursuit (OMP), compressive sampling 
matching pursuit (CoSaMP), and subspace pursuit (SP). 
\end{abstract}
\medskip
\vspace{-2mm}

\pagestyle{empty}

\section{Introduction}
\label{sec1}
Media-based modulation (MBM), a promising modulation scheme for 
wireless communications in multipath fading environments, is 
attracting recent research attention \cite{mbm1}-\cite{mbm6}.
The key features that make MBM different from conventional 
modulation are: $i)$ MBM uses digitally controlled parasitic 
elements external to the transmit antenna that act as radio 
frequency (RF) mirrors to create different channel fade 
realizations which are used as the channel modulation alphabet, 
and $ii)$ it uses indexing of these RF mirrors to convey additional 
information bits. The basic idea behind MBM can be explained as follows.

Placing RF mirrors near a transmit antenna is equivalent to
placing scatterers in the propagation environment close to the
transmitter. The radiation characteristics of each of these scatterers
(i.e., RF mirrors) can be changed by an ON/OFF control
signal applied to it. An RF mirror reflects back the incident
wave originating from the transmit antenna or passes the wave
depending on whether it is OFF or ON. The ON/OFF status of
the mirrors is called as the ‘mirror activation pattern (MAP)’.
The positions of the ON mirrors and OFF mirrors change from
one MAP to the other, i.e., the propagation environment close
to the transmitter changes from one MAP to the other MAP.
Note that in a rich scattering environment, a small perturbation
in the propagation environment will be augmented by many
random reflections resulting in an independent channel. The
RF mirrors create such perturbations by acting as controlled
scatterers, which, in turn, create independent fade realizations
for different MAPs.

If $m_{rf}$ is the number of RF mirrors used, then $2^{m_{rf}}$ MAPs
are possible. If the transmitted signal is received through $n_r$
receive antennas, then the collection of $2^{m_{rf}}$ $n_r$-length
complex channel gain vectors form the MBM channel alphabet. This
channel alphabet can convey $m_{rf}$ information bits through 
MAP indexing. If the antenna transmits a symbol from a conventional 
modulation alphabet denoted by ${\mathbb A}$, then the spectral 
efficiency of MBM is 
$\eta_{{\tiny \mbox{MBM}}}=m_{rf} + \log_2|{\mathbb A}|$ bits per
channel use (bpcu). An implementation of a MBM system consisting of 
14 RF mirrors placed in a compact cylindrical structure with a dipole 
transmit antenna element placed at the center of the cylindrical 
structure has been reported in \cite{mbm4}. Early reporting of the
idea of using parasitic elements for index modulation purposes (in 
the name `aerial modulation') can be found in 
\cite{aerial1},\cite{aerial2}. 

MBM has been shown to possess attractive performance attributes, 
particularly when the number of receive antennas is large 
\cite{mbm1}-\cite{mbm6}. Specifically, MBM with $m_{rf}$ RF mirrors 
and $n_r$ receive antennas over a multipath channel has been shown 
to asymptotically (as $m_{rf}\rightarrow \infty$) achieve the capacity 
of $n_r$ parallel AWGN channels \cite{mbm2}. This suggests that MBM 
can be attractive for use in massive MIMO systems which typically 
employ a large number of receive antennas at the BS. However, the
literature on MBM so far has focused mainly on single-user 
(point-to-point) communication settings. Our first contribution in 
this paper is that, we report MBM in multiuser massive 
MIMO settings and demonstrate significant performance advantages of 
MBM compared to conventional modulation. For example, a bit error 
performance achieved using 500 receive antennas at the BS in a massive 
MIMO system using conventional modulation can be achieved using just 
128 antennas with multiuser MBM. Even multiuser spatial modulation (SM) 
and generalized spatial modulation (GSM) \cite{lajos}-\cite{mugsm} in 
the same massive MIMO settings require more than 200 antennas to achieve 
the same bit error performance. This suggests that multiuser MBM can be 
an attractive scheme for use in the uplink of massive MIMO systems. 

The second contribution relates to exploitation of the inherent sparsity
in multiuser MBM signal vectors for low-complexity signal detection at 
the BS receiver. We resort to compressive sensing (CS) based sparse recovery 
algorithms for this purpose. Several efficient sparse recovery algorithms 
are known in the literature\cite{omp}-\cite{romp}. We propose a multiuser 
MBM signal detection scheme that employs greedy sparse recovery algorithms 
like orthogonal matching pursuit (OMP)\cite{omp}, compressive sampling 
matching pursuit (CoSaMP) \cite{cosamp}, and subspace pursuit 
(SP)\cite{subspace}. Simulation results show that the proposed detection 
scheme using SP achieves very good performance (e.g., significantly 
better performance compared to MMSE detection) at low complexity. This 
demonstrates that CS based sparse signal recovery approach is a natural 
and efficient approach for multiuser MBM signal detection in massive 
MIMO systems.

The rest of the paper is organized as follows. The multiuser MBM system
model is introduced in Sec. \ref{sec2}. The performance of multiuser
MBM with maximum likelihood detection is presented in Sec. \ref{sec3}. 
The proposed sparsity-exploiting detection scheme for multiuser MBM 
signal detection and its performance in massive MIMO systems are 
presented in Sec. \ref{sec4}. Conclusions are presented in Sec. 
\ref{sec5}.

\vspace{-1mm}
\section{Multiuser MBM system model}
\label{sec2}
\begin{figure}[t]
\centering
\includegraphics[width=7.5cm, height=7.75cm]{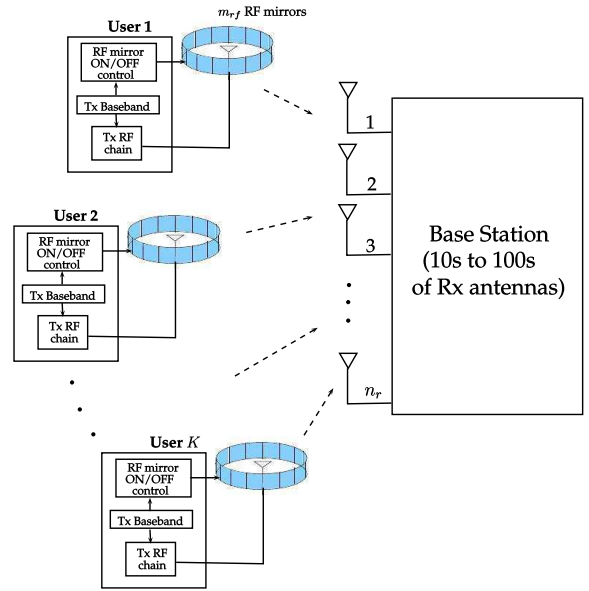}
\caption{Multiuser MBM in a massive MIMO system.}
\label{mbm_mimo}
\vspace{-4mm}
\end{figure}

Consider a massive MIMO system with $K$ uplink users and a BS with $n_r$ 
receive antennas (see Fig. \ref{mbm_mimo}), where $K$ is in the tens (e.g.,
$K=16,32$) and $n_r$ is in the hundreds ($n_r=128,256$). The users employ 
MBM for signal transmission. Each user has a single transmit antenna and 
$m_{rf}$ RF mirrors placed near it. In a given channel use, each user 
selects one of the $2^{m_{rf}}$ mirror activation patterns (MAPs) using 
$m_{rf}$ information bits. A mapping is done between the combinations of 
$m_{rf}$ information bits and the MAPs. An example mapping between 
information bits and MAPs is shown in Table \ref{table1} for $m_{rf}=2$. 
The mapping between the possible MAPs and information bits is made known 
a priori to both transmitter and receiver for encoding and decoding purposes, 
respectively.
\begin{table}[h]
\renewcommand{\arraystretch}{1.3}
\centering
\begin{tabular}{|c|c|c|}
\hline 
Information bits & Mirror 1 status & Mirror 2 status \\ 
\hline 
00 & ON & ON \\ 
\hline 
01 & ON & OFF \\ 
\hline 
10 & OFF & ON \\ 
\hline 
11 & OFF & OFF \\ 
\hline 
\end{tabular}
\vspace{2mm}
\caption{mapping between information bits and MAPs for $m_{rf}=2$.}
\label{table1}
\vspace{-5mm}
\end{table}

Apart from the bits conveyed through the choice of a MAP in a given channel
use as described above, a symbol from a modulation alphabet $\mathbb{A}$ 
(e.g., QAM, PSK) transmitted by the antenna conveys an additional 
$\log_2|\mathbb{A}|$ bits. Therefore, the spectral efficiency of a 
$K$-user MBM system is given by
\begin{equation}
\eta_{{\tiny \mbox{MU-MBM}}}= K(m_{rf}+\log_2{|\mathbb{A}|}) \ \ \mbox{bpcu.}
\end{equation}
For example, a multiuser MBM system with $K=4$, $m_{rf}=2$, and 4-QAM 
has a system spectral efficiency of 16 bpcu. An important point to note
here is that the spectral efficiency per user increases linearly with 
the number of RF mirrors used at each user. 
To introduce the multiuser MBM signal set and the corresponding received
signal vector at the BS, let us first formally introduce the single-user 
MBM signal set. 

\vspace{-3mm}
\subsection{Single-user MBM channel alphabet} 
\label{sec2a}
The MBM channel alphabet of a single user is the set of all channel gain 
vectors corresponding to the various MAPs of that user. Let us define 
$M \Define 2^{m_{rf}}$, where $M$ is the number of possible MAPs 
corresponding to $m_{rf}$ RF mirrors. Let $\mathbf{h}_k^m$ denote the 
$n_r \times 1$ channel gain vector corresponding to the $m$th MAP of 
the $k$th user, where  
${\mathbf h}_{k}^{m}=[h_{1,k}^m \ h_{2,k}^m \ \cdots \ h_{n_r,k}^m]^T$,
$h_{i,k}^m$ is the channel gain corresponding to the $m$th MAP of the 
$k$th user to the $i$th receive antenna, $i=1,\cdots,n_r$, $k=1,\cdots,K$, 
and $m=1,\cdots,M$, and the $h_{i,k}^m$s are assumed to be i.i.d. and 
distributed as $\mathcal{CN}(0,1)$. The MBM channel alphabet for the 
$k$th user, denoted by $\mathbb{H}_k$, is then the collection of these 
channel gain vectors, i.e.,
$\mathbb{H}_k=\{\mathbf{h}_k^1,\mathbf{h}_k^2,\cdots,\mathbf{h}_k^M\}$. 
The MBM channel alphabet of each user is estimated at the BS receiver 
through pilot transmission before data transmission. The number of pilot 
channel uses needed for the estimation of each user's channel alphabet 
grows exponentially in $m_{rf}$. It is also noted that, while the MBM 
channel alphabet of each user needs to be known at the BS receiver for 
detection purposes, the users' transmitters need not know their channel 
alphabets. 

\vspace{-3mm}
\subsection{Single-user MBM signal set} 
\label{sec2b}
Define ${\mathbb A}_0 \Define {\mathbb A}\cup 0$. 
The single-user MBM signal set, denoted by 
$\mathbb{S}_{{\tiny \mbox{SU-MBM}}}$, 
is the set of $M\times 1$-sized MBM signal vectors given by
\begin{align}
\hspace{-4mm}
\mathbb{S}_{{\tiny \mbox{SU-MBM}}} &= \left\{\mathbf{s}_{m,q} \in {\mathbb A}_0^M : m=1,\cdots,M, \ q=1,\cdots,|\mathbb{A}| \right \} \nonumber \\
\mbox{ s.t } \  \mathbf{s}_{m,q} &= [0,\cdots,0,\hspace{-2mm}\underbrace{s_q}_{\mbox{{\scriptsize $m$th coordinate}}}\hspace{-2mm}0,\cdots,0]^T, s_q \in \mathbb{A}, 
\label{ss}
\end{align}
where $m$ is the index of the MAP. That is, an MBM signal vector 
$\mathbf{s}_{m,q}$ in (\ref{ss}) means a complex symbol $s_q \in \mathbb{A}$ 
being transmitted on a channel with an associated channel gain vector 
$\mathbf{h}^m$, where $\mathbf{h}^m$ is the $n_r\times 1$ channel gain 
vector corresponding to the $m$th MAP. Therefore, the $n_r\times 1$ 
received signal vector corresponding to a transmitted MBM signal vector 
$\mathbf{s}_{m,q}$ can be written as 
\begin{equation}
\mathbf{y}=s_q\mathbf{h}^m + \mathbf{n},
\end{equation}
where $\mathbf{n} \in \mathbb{C}^{n_r}$ is the AWGN noise vector with 
$\mathbf{n} \sim \mathcal{CN}(\mathbf{0}, \sigma^2\mathbf{I})$.   
The size of the single-user MBM signal set is 
$|\mathbb{S}_{{\tiny \mbox{SU-MBM}}}|= M|{\mathbb A}|$. For example, 
if $m_{rf}=2$ and $|{\mathbb A}|=2$ (i.e., BPSK ), then 
$|\mathbb{S}_{{\tiny \mbox{SU-MBM}}}|=8$, and the corresponding MBM 
signal set is given by

\vspace{-2mm}
\begin{small}
\begin{equation}
\hspace{-0.0mm}
\mathbb{S}_{{\tiny \mbox{SU-MBM}}}=
\left\{
\begin{bmatrix}
1 \\ 0 \\ 0\\ 0
\end{bmatrix}\hspace{-1.25mm},
\begin{bmatrix}
-1 \\ 0 \\ 0\\ 0
\end{bmatrix}\hspace{-1.25mm},
\begin{bmatrix}
0 \\ 1 \\ 0\\ 0
\end{bmatrix}\hspace{-1.25mm},
\begin{bmatrix}
0 \\ -1 \\ 0\\ 0
\end{bmatrix}\hspace{-1.25mm}, 
\begin{bmatrix}
0 \\ 0 \\ 1\\ 0
\end{bmatrix}\hspace{-1.25mm},
\begin{bmatrix}
0 \\ 0 \\ -1\\ 0
\end{bmatrix}\hspace{-1.25mm},
\begin{bmatrix}
0 \\ 0 \\ 0\\ 1
\end{bmatrix}\hspace{-1.25mm},
\begin{bmatrix}
0 \\ 0 \\ 0\\ -1
\end{bmatrix}\hspace{-1.0mm}
\right \}\hspace{-0.5mm}. \hspace{-4mm}
\label{mbm_sigset}
\end{equation}
\end{small}

\vspace{-2mm}
\subsection{Multiuser MBM received signal}
\label{sec2c}
With the above definitions of single-user MBM channel alphabet and
signal set, the multiuser MBM signal set with $K$ users is given by 
$\mathbb{S}_{{\tiny \mbox{MU-MBM}}}=\mathbb{S}_{{\tiny \mbox{SU-MBM}}}^K$.
Let $\mathbf{x}_k \in \mathbb{S}_{{\tiny \mbox{SU-MBM}}}$ denote the 
transmit MBM signal vector from the $k$th user. Let 
$\mathbf{x} = \left[\mathbf{x}_1^T \ \mathbf{x}_2^T \ \cdots \ \mathbf{x}_K^T \right]^T \in  \mathbb{S}_{{\tiny \mbox{MU-MBM}}}$  
denote the vector comprising of the transmit MBM signal vectors from 
all the $K$ users. Let $\mathbf{H} \in \mathbb{C}^{n_r \times KM}$ denote 
the channel gain matrix given by 
$\mathbf{H}=[\mathbf{H}_1 \ \mathbf{H}_2 \ \cdots \ \mathbf{H}_K]$, where 
$\mathbf{H}_k=[\mathbf{h}_k^1 \ \mathbf{h}_k^2 \ \cdots \ \mathbf{h}_k^M] \in \mathbb{C}^{n_r \times M}$, and 
$\mathbf{h}_k^m$ is the channel gain vector of the $k$th user 
corresponding to $m$th MAP as defined before. The $n_r\times 1$ multiuser 
received signal vector at the BS is then given by
\begin{equation}
\mathbf{y} = \mathbf{Hx} + \mathbf{n},
\label{sys}
\end{equation}
where $\mathbf{n}$ is the $n_r\times 1$ AWGN noise vector with 
$\mathbf{n} \sim \mathcal{CN}(\mathbf{0}, \sigma ^2 \mathbf{I})$. 
 
\section{Performance of multiuser MBM}
\label{sec3}
In this section, we analyze the BER performance of multiuser MBM 
under maximum likelihood (ML) detection. We obtain an upper bound 
on the BER which is tight at moderate to high SNRs. We also present 
a comparison between the BER performance of multiuser MBM and those 
of other multiuser schemes that employ conventional modulation, 
spatial modulation, and generalized spatial modulation.

\vspace{-2mm}
\subsection{Upper bound on BER}
\label{sec3a}
The ML detection rule for the multiuser MBM system model in (\ref{sys}) 
is given by
\begin{equation}
\hat{\mathbf{x}} = \argmin_{\mathbf{x} \in \mathbb{S}_{{\tiny \mbox{MU-MBM}}}} \| \mathbf{y} - \mathbf{Hx} \|^2,
\label{ML}
\end{equation}
which can be written as 
\begin{equation}
\hat{\mathbf{x}}=\argmin_{\mathbf{x} \in  \mathbb{S}_{{\tiny \mbox{MU-MBM}}}}\left( \Vert\mathbf{H}\mathbf{x}\Vert^2-2\mathbf{y}^T\mathbf{H}\mathbf{x}\right) .
\end{equation}
The  pairwise error probability (PEP) that the receiver decides in favor 
of the signal vector $\mathbf{x}_2$ when $\mathbf{x}_1$  was transmitted, 
given the channel matrix $\mathbf{H}$ can be written as
\begin{eqnarray}
\label{PEP}
PEP & = & P(\mathbf{x}_1 \rightarrow \mathbf{x}_2| \mathbf{H}) \nonumber \\
& = & P\left(2\mathbf{y}^T\mathbf{H}(\mathbf{x}_2-\mathbf{x}_1)>(\Vert\mathbf{H}\mathbf{x}_2\Vert^2-\Vert\mathbf{H}\mathbf{x}_1\Vert^2) | \mathbf{H}\right) \nonumber \\
& = & P\left(2\mathbf{n}^T\mathbf{H}(\mathbf{x}_2-\mathbf{x}_1)>\Vert\mathbf{H}(\mathbf{x}_2-\mathbf{x}_1)\Vert^2 |\mathbf{H}\right).
\end{eqnarray}
Defining $z\triangleq 2\mathbf{n}^T\mathbf{H}(\mathbf{x}_2-\mathbf{x}_1)$, 
we observe that $z \sim \mathcal{N} \left(0, 2\sigma^2\Vert\mathbf{H}(\mathbf{x}_2-\mathbf{x}_1)\Vert^2 \right)$.
Therefore, we can write 
\begin{equation}
P(\mathbf{x}_1 \rightarrow \mathbf{x}_2| \mathbf{H})=Q\left(\frac{\Vert \mathbf{H}(\mathbf{x}_2-\mathbf{x}_1)\Vert}{\sqrt{2}\sigma}\right),
\label{pep1}
\end{equation}
where $Q(x)=\frac{1}{\sqrt{2\pi}}\int_{x}^{\infty}e^{\frac{-t^2}{2}}dt$.
The conditional PEP expression in \eqref{pep1} can be written as
\begin{equation}
P(\mathbf{x}_1 \hspace{-0.5mm} \rightarrow \hspace{-0.5mm} \mathbf{x}_2 | \mathbf{H}) =
Q\hspace{-0.5mm}\left(\hspace{-1.5mm} \sqrt{ \frac{1}{2\sigma^2}\left \| \sum_{l=1}^{KM} (x_{1,l} - x_{2,l})\mathbf{h}_l \right \|^2}\right),
\label{pep2}
\end{equation}
where $x_{1,l}$ and $x_{2,l}$ are $l$th entries of $\mathbf{x}_1$ and 
$\mathbf{x}_2$, respectively, and $\mathbf{h}_l$ is the $l$th column of 
$\mathbf{H}$. The argument of $Q(\cdot)$ in \eqref{pep2} has the central 
$\chi ^2$-distribution with $2n_r$ degrees of freedom. The computation of 
the unconditional PEPs requires the expectation of $Q(\cdot)$ with respect 
to $\mathbf{H}$, which can be obtained as follows\cite{erroranalysis}:
\begin{align}
\hspace{-3mm}
P(\mathbf{x}_1 \rightarrow \mathbf{x}_2) &= \mathbb{E}_\mathbf{H} \left[ P(\mathbf{x}_1 \rightarrow \mathbf{x}_2 | \mathbf{H}) \right] \nonumber \\
&=f(\alpha)^{n_r} \sum_{i=0}^{n_r-1}  \binom{n_r-1+i}{i} (1-f(\alpha))^i,  
\end{align}
where 
$f(\alpha) \triangleq \dfrac{1}{2} \left(1-\sqrt{\dfrac{\alpha}{1+\alpha}} \right)$,
$\alpha \triangleq \dfrac{1}{4\sigma^2} \sum\limits_{l=1}^{KM} \theta _l$, 
and $\theta _l \triangleq | x_{1,l} - x_{2,l} |^2$. 
Now, an upper bound on the bit error probability using union bound can be 
obtained as
\begin{equation}
P_e \leq \dfrac{1}{2^{\eta_{{\tiny \mbox{MU-MBM}}}}} \hspace{-2mm} \sum_{\mathbf{x}_1 \in \mathbb{S}_{\tiny \mbox{MU-MBM}}} \sum_{\mathbf{x}_2 \in \mathbb{S}_{\tiny \mbox{MU-MBM}} \setminus \mathbf{x}_1} \hspace{-4mm} P(\mathbf{x}_1 \rightarrow \mathbf{x}_2) \dfrac{d_H(\mathbf{x}_1, \mathbf{x}_2)}{\eta_{{\tiny \mbox{MU-MBM}}}},
\label{ber} 
 \end{equation}
where $d_H(\mathbf{x}_1, \mathbf{x}_2)$ is the Hamming distance between 
the bit mappings corresponding to $\mathbf{x}_1$ and $\mathbf{x}_2$.

\vspace{-2mm}
\subsection{Numerical results}
\label{sec3b}
We evaluated the BER performance of multiuser MBM (MU-MBM) using the BER 
upper bound derived above as well as simulations. For the purpose of 
initial comparisons with other systems, we consider a MU-MBM system 
with $K=2$, $n_r=8$, $m_{rf}=3$, BPSK, and 4 bpcu per user. Let $n_t$ 
and $n_{rf}$ denote the number transmit antennas and transmit 
RF chains, respectively, at each user. Note that in the considered 
MU-MBM system, each user uses one transmit antenna and one transmit 
RF chain, i.e., $n_t=n_{rf}=1$. We compare the performance of the 
above MU-MBM system with those of three other multiuser systems which 
use $i)$ conventional modulation (CM), $ii)$ spatial modulation (SM), 
and $iii)$ generalized spatial modulation (GSM). The multiuser system 
with conventional modulation (MU-CM) uses $n_t=n_{rf}=1$ at each user 
and employs 16-QAM to achieve the same spectral efficiency of 4 bpcu 
per user. The multiuser system with SM (MU-SM) uses $n_t=2$, $n_{rf}=1$, 
and 8-QAM, achieving a spectral efficiency of $\log_2n_t+\log_2|\mathbb A|= 
\log_22+\log_28=4$ bpcu per user. The multiuser system with GSM (MU-GSM) 
uses $n_t=4$, $n_{rf}=2$, and BPSK, achieving a spectral efficiency of 
$\lfloor\log_2\binom{n_t}{n_{rf}}\rfloor+\log_2|\mathbb A|= 
\lfloor\log_2\binom{4}{2}\rfloor+\log_22= 4$ bpcu per user.

\begin{figure}
\centering
\includegraphics[width=9.0cm, height=6.25cm]{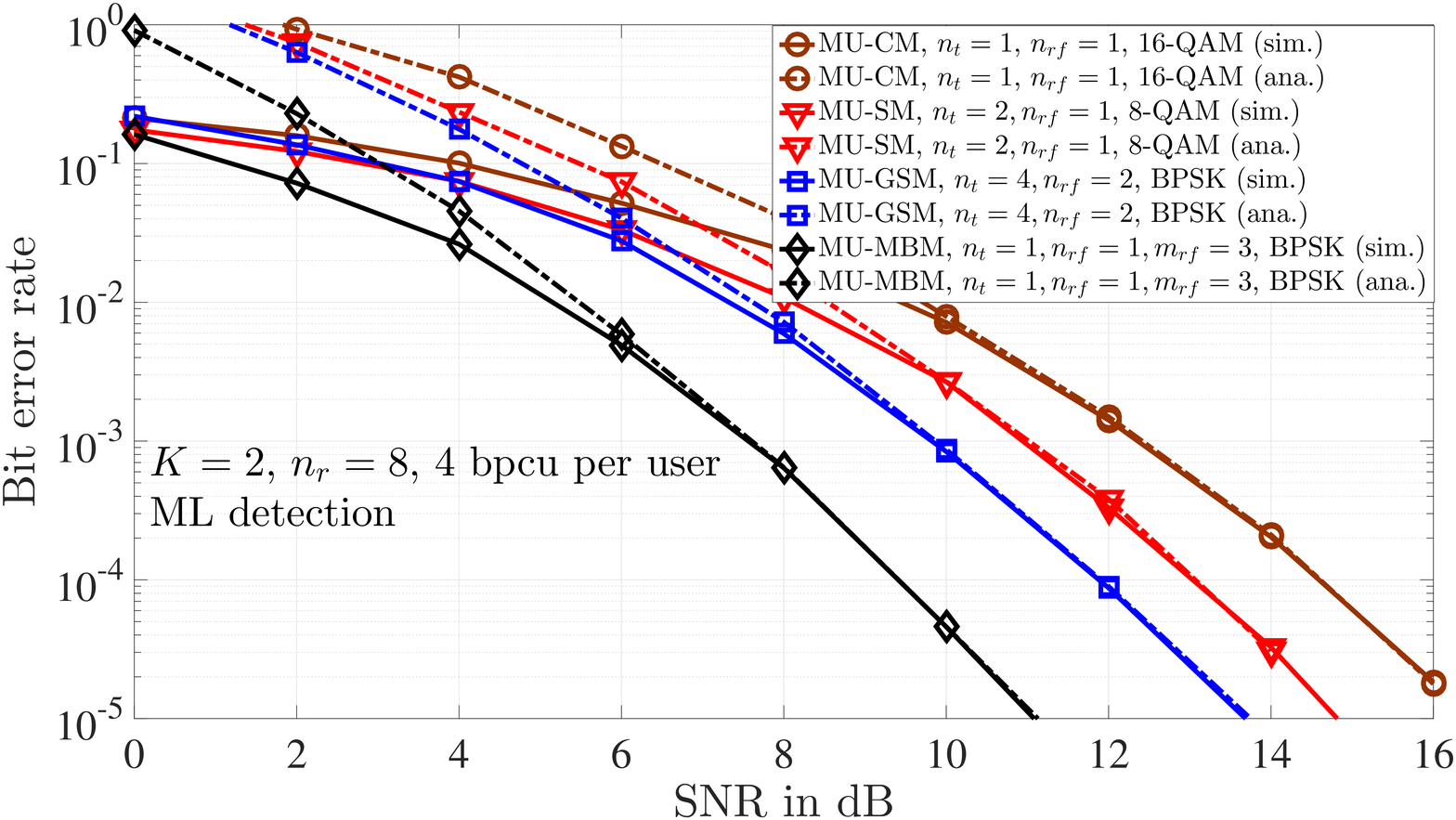}
\vspace{-5mm}
\caption{BER performance of MU-MBM, MU-CM, MU-SM, and MU-GSM with 
$K=2$, $n_r=8$, 4 bpcu per user, and ML detection. Analysis and
simulations.}
\label{ml}
\vspace{-4mm}
\end{figure}

Figure \ref{ml} shows the BER performance of the MU-MBM, MU-CM, MU-SM, 
and MU-GSM systems described above. First, it can be 
observed that the analytical upper bound is very tight at moderate 
to high SNRs. Next, in terms of performance comparison between the 
considered systems, the following  inferences can be drawn from 
Fig. \ref{ml}.
\begin{itemize}
\item 	The MU-MBM system achieves the best performance among all the 
	four systems considered. For example, MU-MBM performs better 
	by about 5 dB, 4 dB, 2.5 dB compared to MU-CM, MU-SM, and
	MU-GSM systems, respectively, at a BER of $10^{-5}$. 
\item 	The better performance of MU-MBM can be 
	attributed to more bits being conveyed through mirror indexing,  
	which allows MU-MBM to use lower-order modulation alphabets (BPSK) 
	compared to other systems which may need higher-order alphabets 
	(8-QAM, 16-QAM) to achieve the same spectral efficiency. 
\item	MU-MBM performs better than MU-GSM though both use BPSK in this
	example. This can be attributed to the good distance properties 
	of the MBM signal set \cite{mbm2}. 
\end{itemize}

Note that though the results in Fig. \ref{ml} illustrate the performance 
superiority of MU-MBM over MU-CM, MU-SM, and MU-GSM, they are presented 
only for a small system with $K=2$ and $n_r=8$. This is because ML 
detection is prohibitively complex for systems with large $K$ and $n_r$ 
(ML detection is exponentially complex in $K$). 
However, massive MIMO systems are characterized by $K$ in the tens and
$n_r$ in the hundreds. Therefore, low-complexity detection schemes which
scale well for such large-scale MU-MBM systems are needed. To address 
this need, we resort to exploiting the inherent sparse nature of the 
MBM signal vectors, and devise a compressive sensing based detection 
algorithm in the following section. 

\vspace{-1mm}
\section{Sparsity-exploiting detection of multiuser MBM signals} 
\label{sec4}
It is evident from the example signal set in \eqref{mbm_sigset} that the 
MBM signal vectors are inherently sparse. An MBM signal vector has only 
one non-zero element out of $M$ elements, leading to a sparsity factor of 
$1/M$. For example, consider an MBM signal set with $m_{rf}=4$ and 
$M=2^{m_{rf}}=16$. Out of 16 elements in a signal vector, only one 
element is non-zero resulting in a sparsity factor of $1/16$. 
Exploitation of this inherent sparsity to devise detection algorithms 
can lead to efficient signal detection at low complexities. Accordingly, 
we propose a low-complexity MU-MBM signal detection scheme that employs 
compressive sensing based sparse reconstruction algorithms like OMP, 
CoSaMP, and SP.       
 
\vspace{-3mm}
\subsection{Proposed sparsity-exploiting detection algorithm}
\label{sec4a}
We first model the MU-MBM signal detection problem as a sparse 
reconstruction problem and then employ greedy algorithms for signal 
detection. Sparse reconstruction is concerned with finding an approximate 
solution to the following problem \cite{sparse}:
\begin{equation}
\min\limits_{\mathbf{x}} \| \mathbf{x} \|_0 \mbox{ subject to } \mathbf{y} = \mathbf{\Phi x + n},
\label{sparse}
\end{equation}
where $\mathbf{\Phi} \in \mathbb{C}^{m \times n}$ is called the 
measurement matrix, $\mathbf{x} \in \mathbb{C}^{n}$ is the complex 
input signal vector, $\mathbf{y} \in \mathbb{C}^{m}$ is the noisy 
observation corresponding to the input signal, and 
$\mathbf{n} \in \mathbb{C}^m$ is the complex noise vector. The 
MU-MBM signal detection problem at the BS in (\ref{sys}) can be 
modeled as a sparse recovery problem in \eqref{sparse}, with the 
measurement matrix being the channel matrix 
$\mathbf{H} \in \mathbb{C}^{n_r \times KM}$, the noisy observation 
being the received signal vector $\mathbf{y} \in \mathbf{C}^{n_r}$, 
and the input being the MU-MBM transmit signal vector 
$\mathbf{x} \in \mathbb{S}_{{\tiny \mbox{MU-MBM}}} $. The noise 
vector is additive complex Gaussian with 
$\mathbf{n} \sim \mathcal{CN}(\mathbf{0}, \sigma^2 \mathbf{I})$. 

Greedy algorithms achieve sparse reconstruction in an iterative 
manner. They decompose the problem of sparse recovery into a two 
step process; recover the support of the sparse vector first,
and then obtain the non-zero values over this support. For example, 
OMP starts with an initial empty support set, an initial solution  
$\mathbf{x}^0=\mathbf{0}$, and an initial residue 
$\mathbf{r}^0 =\mathbf{y}-\mathbf{\bf Hx}^0 =\mathbf{y}$. In each step, 
OMP updates one coordinate  of the vector $\mathbf{x}$ based on the 
correlation values between the residue vector and the columns of the 
$\mathbf{\bf H}$ matrix. In the $k$th iteration, an element $j_0$ 
given by 
\[
j_0 = \argmax\limits_{j \notin \mathcal{S}^{k-1}} \frac{\mathbf{\bf h}_j^T \mathbf{r}^{k-1}}{\| \mathbf{\bf h}_j \|_2^2}
\]
is added to the support set, where $\mathbf{\bf h}_j$ is the $j$th 
column of $\mathbf{\bf h}$, and $\mathcal{S}^{k-1}$ and $\mathbf{r}^{k-1}$ 
are the support set and residue after $k-1$ iterations, respectively. 
The entries of $\mathbf{x}$ corresponding to the obtained support set are 
computed using least squares. This process is iterated till the stopping 
criteria is met. The stopping criteria can be either a specified error 
threshold or a specified level of sparsity.  

In the SP algorithm, instead of updating one coordinate of $\mathbf{x}$ 
at a time as in OMP, $K$ coordinates are updated at once. The major 
difference between OMP and SP is the following. In OMP, the support set 
is generated sequentially. It starts with an empty set and adds one 
element in every iteration to the existing support set. An element added 
to the support set can not be removed until the algorithm terminates. 
In contrast, SP provides flexibility of refining the support set in 
every iteration. CoSaMP is  similar to SP except that it updates $2K$ 
coordinates in each iteration to the support set instead of updating 
$K$ coordinates as in SP. CoSaMP and SP have superior reconstruction 
capability comparable to convex relaxation methods 
\cite{cosamp},\cite{subspace}. {\bf Algorithm 1} shows the listing of 
the pseudo-code of the proposed sparsity-exploiting detection algorithm 
for MU-MBM signals. 

\begin{algorithm}
\caption{Proposed sparsity-exploiting algorithm for MU-MBM signal 
detection}
\begin{algorithmic}[1]
\State Inputs: $\mathbf{y}, \mathbf{H}, K $  	
\State Initialize: $j=0$
\State  \textbf{repeat}
\State \hspace*{0.4 cm} $\mathbf{\hat{x}}_r=\mbox{SR}(\mathbf{y},\mathbf{H},K+j)$ \Comment{Sparse Recovery algorithm}
\State \hspace*{0.4 cm} $\mathbf{u}^j= \mbox{UAP}(\hat{\mathbf{x}}_r)$ \Comment{Extract User Activity Pattern}
\State \hspace*{0.4 cm} \textbf{if}  $ \| \mathbf{u}^j \| _0 = K $  
\State \hspace*{0.8 cm} \textbf{for} $k= 1 \mbox{ to } K$  
\State  \hspace*{1.2 cm}$ \hat{\mathbf{x}}^k = \argmin\limits_{\mathbf{s} \in \mathbb{S}_{{\tiny \mbox{SU-MBM}}}} \| \hat{\mathbf{x}}_r^k-\mathbf{s} \|^2$  \Comment{Nearest MBM signal \\ \vspace{-3mm} \hspace{5.10cm} mapping}
\State \hspace*{0.8 cm} \textbf{end for}
\State \hspace*{0.8 cm}\textbf{break};
\State \hspace*{0.4 cm}\textbf{else} $j=j+1$ 
\State \hspace*{0.4 cm}\textbf{end if}
\State \textbf{until} $j < K(M-1)$
\State Output: The estimated MU-MBM signal vector 
$$\hat{\mathbf{x}} = [\hat{\mathbf{x}}^{\scriptsize{1}^T}, \hat{\mathbf{x}}^{\scriptsize{2}^T}, \cdots ,\hat{\mathbf{x}}^{\scriptsize{K}^T} ]^T$$
\end{algorithmic}
\label{alg1}
\end{algorithm}

SR in {\bf Algorithm 1} denotes the sparse recovery algorithm, which 
can be any one of OMP, CoSaMP, and SP. The signal vector reconstructed by 
the sparse recovery algorithm is denoted by $\hat{\mathbf{x}}_r$.
Detecting the MU-MBM signal vector involves detecting the MBM signal 
vector transmitted by each user. An MBM signal vector from a user has 
exactly one non-zero entry out of $M$ entries as observed in the 
example MBM signal set in \eqref{mbm_sigset}. Hence, SR is expected to 
reconstruct a MU-MBM signal vector such that the MBM signal sub-vector
corresponding to a given user has only one non-zero entry. But this 
constraint on the expected support set is not built in the general sparse 
recovery algorithms. In general, a sparse recovery algorithm can output 
$K$ non-zero elements at any of the $KM$ locations of $\hat{\mathbf{x}}_r$. 
To overcome this issue, we define user activity pattern (UAP), denoted by 
$\mathbf{u}$, as a $K$-length vector with $k$th entry as $\mathbf{u}_k=1$ 
if there is at least one non-zero entry in the $k$th user's recovered MBM 
signal vector, and $\mathbf{u}_k=0$ otherwise. A valid reconstructed 
signal vector is one which has all ones in $\mathbf{u}$. SR is used 
multiple times with a range of sparsity estimates starting from $K$ 
($K+j$ in the algorithm listing) till the valid UAP is obtained (i.e., 
till the algorithm reconstructs at least one non-zero entry for each 
user's MBM signal vector). 

In the algorithm listing, $\mathbf{u}^j$ denotes the UAP at the $j$th 
iteration. On recovering an $\hat{\mathbf{x}}_r$ with valid UAP, the 
MBM signal vector of each user is mapped to the nearest (in the Euclidean 
sense) MBM signal vector in $\mathbb{S}_{{\tiny \mbox{SU-MBM}}}$. This is 
shown in the Step 8 in the algorithm listing, where $\hat{\mathbf{x}}_r^k$ 
denotes the recovered MBM signal vector of the $k$th user and 
$\hat{\mathbf{x}}^k$ denotes the MBM signal vector to which 
$\hat{\mathbf{x}}_r^k$ gets mapped to. Finally, the MU-MBM signal 
vector is obtained by concatenating the detected MBM signal vectors 
of all the users, i.e., , 
$\hat{\mathbf{x}} = [\hat{\mathbf{x}}^{\scriptsize{1}^T}, \hat{\mathbf{x}}^{\scriptsize{2}^T}, \cdots ,\hat{\mathbf{x}}^{\scriptsize{K}^T} ]^T$.

The decoding of information bits from the detected MBM signal vector 
of a given user involves decoding of mirror index bits and QAM symbol
bits of that user. The mirror index bits are decoded from the MAP of 
the detected MBM signal vector and the QAM bits are decoded from the 
detected QAM symbol.

\vspace{-1mm}
\subsection{Performance results in massive MIMO system}
\label{sec4b}
In this subsection, we present the BER performance of MU-MBM systems 
in a massive MIMO setting (i.e., $K$ in the tens and $n_r$ in the
hundreds) when the proposed {\bf Algorithm 1} is used for MU-MBM signal 
detection at the BS. In the same massive MIMO setting, we evaluate
the performance of other systems that use conventional modulation 
(MU-CM), spatial modulation (MU-SM), and generalized spatial modulation 
(MU-GSM), and compare them with the performance achieved by MU-MBM. 
The proposed {\bf Algorithm 1} is also used for the detection of 
MU-SM and MU-GSM. It is noted that the MU-SM and MU-GSM signal
vectors are also sparse to some extent; the sparsity factors in 
MU-SM and MU-GSM are $1/n_t$ and $n_{rf}/n_t$, respectively. So the 
use of the proposed algorithm for detection of these signals is also 
appropriate. ML detection is used to detect MU-CM signals (this is 
possible for MU-CM with sphere decoding for $K=16$, i.e., 32 real 
dimensions). 

\begin{figure}
\centering
\includegraphics[width=9cm, height=6.25cm]{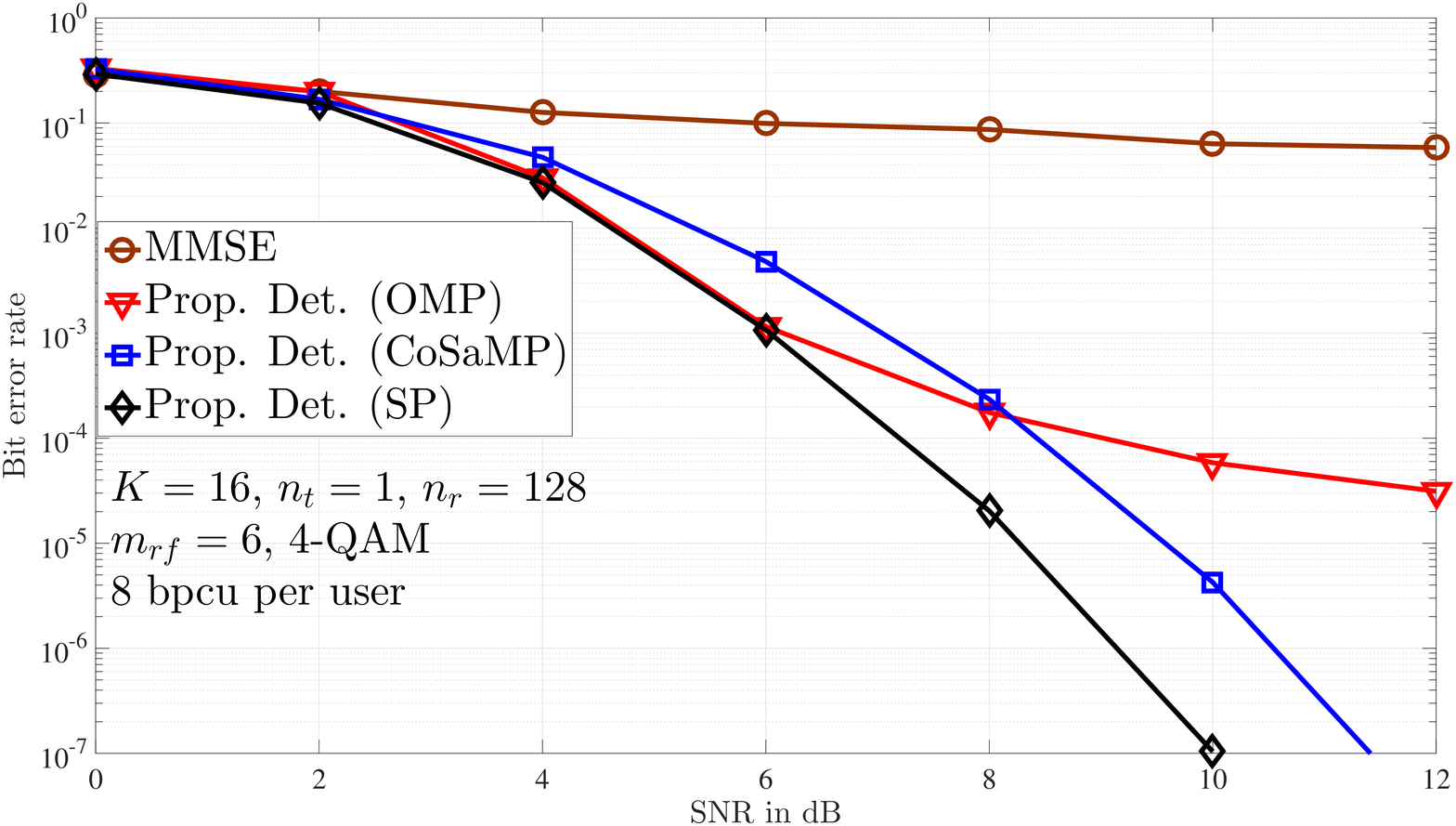}
\vspace{-6mm}
\caption{BER performance of MU-MBM in a massive MIMO system with $K=16$, 
$n_r=128$, $n_t=1$, $n_{rf}=1$, $m_{rf}=6$, 4-QAM, 8bpcu per user, using 
the proposed detection algorithm. MMSE detection performance is also shown 
for comparison.}
\label{fig:sparsealgos}
\vspace{-2mm}
\end{figure}

{\em MU-MBM performance using proposed algorithm:}
Figure \ref{fig:sparsealgos} shows the performance of MU-MBM system 
using the proposed algorithm with $i)$ OMP, $ii)$ CoSaMP, and 
$iii)$ SP. MMSE detection performance is also shown for comparison.
A massive MIMO system with $K=16$ and $n_r=128$ is considered.
Each user uses $n_t=1$, $n_{rf}=1$, $m_{rf}=6$, and 4-QAM. This 
results in a spectral efficiency of 8 bpcu per user, and a 
sparsity factor of $1/64$. From Fig. \ref{fig:sparsealgos}, we 
observe that the proposed algorithm with OMP, CoSaMP, and SP 
achieve significantly better performance compared to MMSE.  
Among the the use of OMP, CoSaMP, and SP in the proposed algorithm,
use of SP gives the best performance. This illustrates the superior
reconstruction/detection advantage of the proposed algorithm with
SP. We will use the proposed algorithm with SP in the subsequent
performance results figures. It is noted that the complexity of proposed 
algorithm is also quite favorable; the complexity of the proposed 
algorithm with SP and that of MMSE are $O(K^2Mn_r)$ and $O(K^3M^3)$, 
respectively.  

\begin{figure}
\centering
\includegraphics[width=9cm, height=6.25cm]{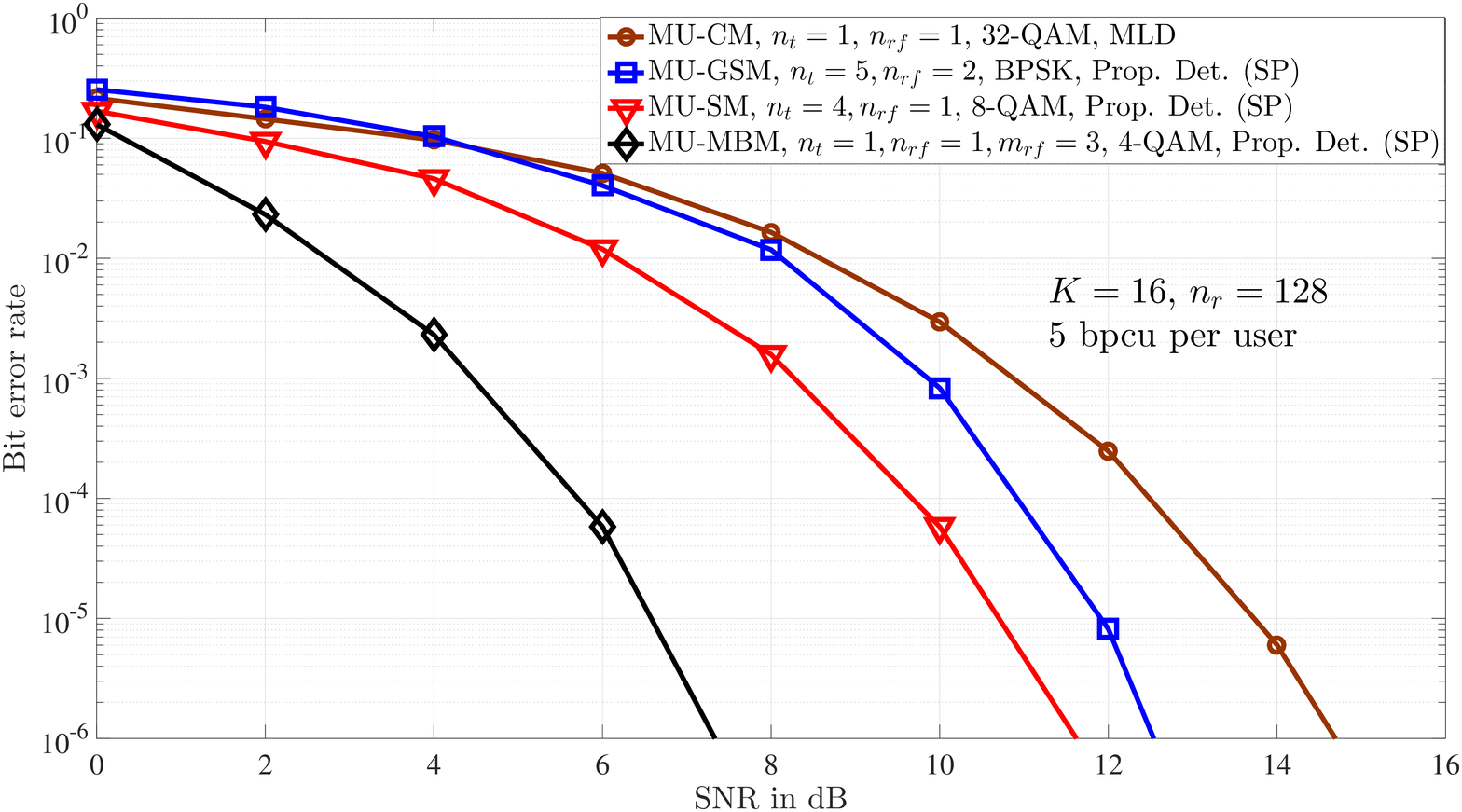}
\vspace{-6mm}
\caption{BER performance MU-MBM, MU-CM, MU-SM, and MU-GSM in a 
massive MIMO setting with $K=16$, $n_r=128$, and 5 bpcu per user.}
\vspace{-2mm}
\label{fig:spdetection}
\end{figure}

{\em Performance of MU-MBM, MU-SM, MU-GSM:} 
Figure \ref{fig:spdetection} shows a BER performance comparison between
MU-MBM, MU-CM, MU-SM, and MU-GSM in a massive MIMO setting with $K=16$ 
and $n_r=128$. The proposed algorithm with SP is used for detection 
in MU-MBM, MU-SM, and MU-GSM. ML detection is used for MU-CM. 
The spectral efficiency is fixed at 5 bpcu per user for all the four
schemes. MU-MBM achieves this spectral efficiency with $n_t=1$, $n_{rf}=1$,
$m_{rf}=3$, and 4-QAM. MU-CM uses $n_t=1$, $n_{rf}=1$, and 32-QAM to
achieve 5 bpcu per user. To achieve the same 5 bpcu per user, MU-SM 
uses $n_t=4$, $n_{rf}=1$, and 8-QAM, and MU-GSM uses $n_t=5$, $n_{rf}=2$,
and BPSK. The sparsity factors in MU-MBM, MU-SM, and MU-GSM are 
$1/8$, $1/4$, and $2/5$, respectively. It can be seen that, MU-MBM 
clearly outperforms MU-CM, MU-SM, and MU-GSM. For example, at a BER
of $10^{-5}$, MU-MBM outperforms MU-CM, MU-GSM, and MU-SM by about 
7 dB, 5 dB, and 4 dB, respectively. The performance advantage of
MU-MBM can be mainly attributed to its better signal distance 
properties \cite{mbm2}. MU-MBM is also benefited by its lower 
sparsity factor as well as the possibility of using lower-order 
QAM size because of additional bits being conveyed through indexing 
mirrors. 

\begin{figure}
\centering
\includegraphics[width=9cm, height=6.25cm]{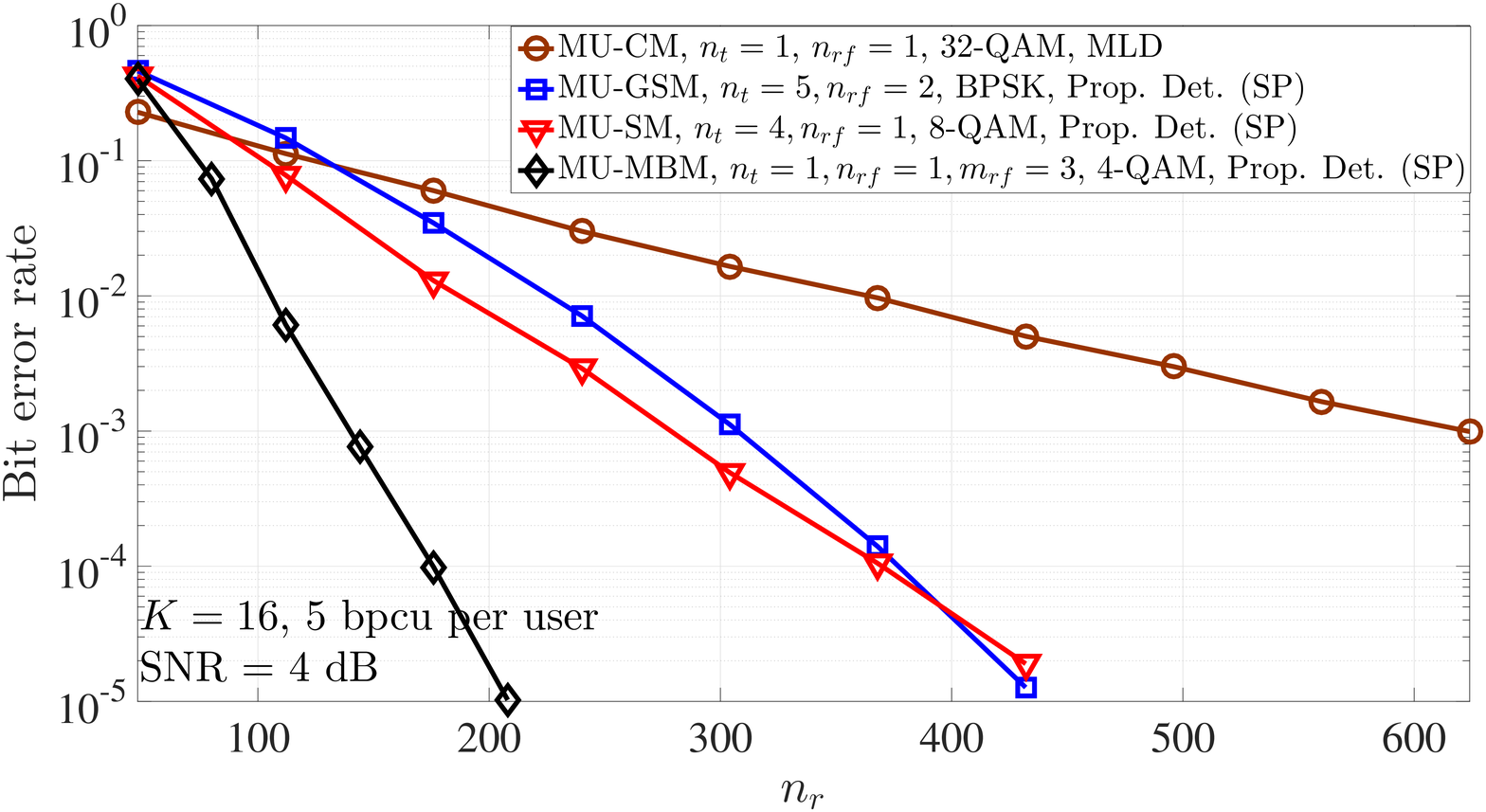}
\vspace{-6mm}
\caption{BER performance MU-MBM, MU-CM, MU-SM, and MU-GSM as a function
of $n_r$ in a massive MIMO setting with $K=16$, 5 bpcu per user, and 
SNR = 4 dB.}
\label{fig:effectofnr}
\vspace{-2mm}
\end{figure}

{\em Effect of number of BS receive antennas}: 
Figure \ref{fig:effectofnr} shows an interesting result which 
demonstrates MU-MBM's increasing performance gain compared to 
MU-CM, MU-SM, and MU-GSM as the number of BS receive antennas
is increased. A massive MIMO system with $K=16$ and 5 bpcu per
user is considered. The parameters of the four schemes are the
same as those in Fig. \ref{fig:spdetection} except that here
SNR is fixed at 4 dB and $n_r$ is varied from 48 to 624. It is 
interesting to observe that a performance that could be achieved 
using 500 antennas at the BS in a massive MIMO system that uses 
conventional modulation ($3\times 10^{-3}$ BER for MU-CM at 
$n_r=500$ with ML detection) can be achieved using just 128 
antennas when MU-MBM is used ($3\times 10^{-3}$ BER for MU-MBM 
at $n_r=128$ with proposed detection). MU-SM and MU-GSM
also achieve better performance compared to MU-CM, but they
too require more than 200 antennas to achieve the same BER.
This increasing performance advantage of MU-MBM for increasing $n_r$ 
can be mainly attributed to its better signal distance properties 
particularly when $n_r$ is large \cite{mbm2}. This indicates that 
multiuser MBM can be a very good scheme for use in the uplink of 
massive MIMO systems. 
   
\vspace{-1mm}
\section{Conclusions}
\label{sec5}
We investigated the use of media-based modulation (MBM), a recent and 
attractive modulation scheme that employs RF mirrors (parasitic elements) 
to convey additional information bits through indexing of these mirrors, 
in massive MIMO systems. Our results demonstrated significant performance 
advantages possible in multiuser MBM compared to multiuser schemes that 
employ conventional modulation, spatial modulation, and generalized 
spatial modulation. Motivated by the possibility 
of exploiting the inherent sparsity in multiuser MBM signal vectors, 
we proposed a detection scheme based on compressive sensing algorithms 
like OMP, CoSaMP, and subspace pursuit. The proposed detection scheme 
was shown to achieve very good performance (e.g., significantly better 
performance compared to MMSE detection) at low complexity, making it 
suited for multiuser MBM signal detection in massive MIMO systems.
Channel estimation, effect of imperfect knowledge of the channel 
alphabet at the receiver, and effect of spatial correlation are 
interesting topics for further investigation.

\end{document}